\documentclass[10pt,final,journal,twocolumn,letterpaper]{IEEEtran}
\usepackage{times}                               
\usepackage{verbatim}                            
\usepackage{graphicx}
\usepackage{tabularx}
\usepackage{algo}
\usepackage{url}
\usepackage{amsmath}
\usepackage{amssymb}
\usepackage{amsfonts}
\usepackage{textcomp,xspace}
\usepackage{enumerate}
\usepackage[ruled, vlined]{algorithm2e}
\usepackage{multirow}
\usepackage{subfigure}
\usepackage[usenames,dvips]{color}
\usepackage[table]{xcolor}
\usepackage{authblk}
\definecolor{lightgray}{gray}{0.9}

\begin{document}
\pagestyle{plain}

\title{Ancilla-Quantum Cost Trade-off during Reversible Logic Synthesis using Exclusive Sum-of-Products}

\author[$\dagger$]{Anupam Chattopadhyay}
\author[$\star$]{Nilanjan Pal\thanks{This work was accomplished during the visit of the second author to MPSoC Architectures Research Group, RWTH Aachen, funded by DAAD WISE fellowship.}}
\author[$\dagger$]{Soumajit Majumder}

\affil[$\dagger$]{MPSoC Architectures Research Group, RWTH Aachen University, Aachen 52074, Germany}
\affil[$\star$]{IIT Kharagpur, Kharagpur, West Bengal, India}

\maketitle

\begin{abstract}
Emerging technologies with asymptotic zero power dissipation, such as quantum computing, require the logical operations to be done in a reversible manner. 
In recent years, the problem of synthesizing Boolean functions in the reversible logic domain has gained significant research attention. The efficiency of 
the synthesis methods is measured in terms of quantum cost, gate cost, garbage lines, logic depth and speed of synthesis. In this paper, we present a 
modification of the existing approaches based on Exclusive Sum-of-Products (ESOP), which allows to explore the trade-off between quantum cost and garbage lines. 
The proposed technique adds a new dimension to the reversible logic synthesis solutions. We demonstrate by detailed experiments that controlled improvement 
in quantum cost and gate count by increasing garbage count can be achieved. In some cases, improved quantum cost and gate count compared to state-of-the-art 
synthesis methods are reported. Furthermore, we propose a novel rule-based approach to achieve ancilla-free reversible logic synthesis starting from an 
ESOP formulation.
\end{abstract}


\section{Introduction}
\begin{sloppypar}
Charles Bennett, in 1973, showed that reversible computation at logical plane can to be done in reversible manner to achieve theoretically zero power 
dissipation~\cite{bennett} in physical reversible computing. Several target technologies in nanoscale computing as well as technologies beyond CMOS, in 
particular Quantum computing, relies on reversible logic. Recent studies show that reversible logic synthesis can be applied to other directions~\cite{wille_encoder} 
as well. Synthesis of a given Boolean function for reversible logic is, therefore, an important open problem. Before further discussion, we present some 
background on reversible logic.
\end{sloppypar}

\subsection{Preliminaries}
\begin{sloppypar}
An $n$-variable Boolean function $f$ is a mapping $f : \{0, 1\}^n \rightarrow \{0, 1\}$, which can also be represented by a truth table. Alternatively, an 
$n$-variable Boolean function $f(x_1, \ldots, x_n)$ can be considered to be a multivariate polynomial over $GF(2)$. This polynomial can be expressed as a 
sum-of-products representation of all distinct $k$-th order products $(0 \leq k \leq n)$ of the variables. This representation of $f$ is called the {\em 
algebraic normal form} (ANF) of $f$. The number of variables in the highest order product term with nonzero coefficient is called the {\em algebraic degree}, 
or simply the degree of $f$ and denoted by $deg(f)$. The ANF representation is also known as Positive-Polarity Reed-Muller expression (PPRM). This is a canonical 
form of a more general Exclusive Sum-Of-Product (ESOP) realization.
\end{sloppypar}

\begin{sloppypar}
An $n$-variable Boolean function is \textit{reversible} if all its output patterns map uniquely to an input pattern and vice-versa. It can be expressed as an $n$-input, 
$n$-output bijection or alternatively, as a permutation function over the truth value set $\{0, 1, \ldots 2^{n-1}\}$. The problem of reversible logic synthesis is to map 
such a reversible Boolean function on a reversible logic gate library.
\end{sloppypar}

\begin{sloppypar}
The gates are characterized by their implementation cost in quantum technologies, which is dubbed as Quantum Cost (QC)~\cite{miller_ismvl11,maslov_benchmark,barenco}.
Few prominent classical reversible logic gates are presented below.
\end{sloppypar}

\begin{itemize}
\item NOT gate: $f(A)$ = $\overline{A}$.
\item CNOT gate: $f(A,B)$ = $(A,A \oplus B)$.
\item CCNOT gate: Also known as Toffoli gate. $f(A,B,C)$ = $(A,B,AB \oplus C)$. This gate can be generalized with $Tof_n$ gate, where first $n-1$ 
variables are used as control lines. NOT and CNOT gates are denoted as $Tof_1$ and $Tof_2$ respectively.
\item Peres gate: A sequence of $Tof_3(A, B, C)$, $Tof_2(A, B)$ or its inverse is known as Peres gate.
\item Controlled Swap gate: Also known as Fredkin gate. $f(A,B,C)$ = $(A, \overline{A}.B + A.C, \overline{A}.C + A.B)$. This gate can be 
generalized with $Fred_n$ gate $(n > 1)$, where first $n-2$ variables are used as control lines.
\end{itemize}

\begin{sloppypar}
Multiple sets of reversible gates form an universal gate library for realizing classical Boolean functions such as, (i) NCT: NOT, CNOT, Toffoli. (ii) NCTSF: NOT, CNOT, 
Toffoli, SWAP, Fredkin. (iii) GT: $Tof_n$. (iv) GTGF: $Tof_n$ and $Fred_n$.
\end{sloppypar}

\begin{sloppypar}
Following the QC assumption of~\cite{barenco}, we use a QC of $1$ for all $2$-qubit elementary reversible logic gates. Optimized implementation of larger gates are 
assumed for the corresponding QC computation, following~\cite{maslov_benchmark} and~\cite{kerntopf_gen_peres}. The used QC values of a generalized Toffoli gate ($Tof_{n+1}$) is $2n^2 - 2n + 1$, of a generalized Fredkin gate ($Fred_{n+1}$) is $2n^2 - 2n + 3$, of a generalized Peres gate ($Per_n$) 
is $n^2$.
\end{sloppypar}

\begin{sloppypar}
Reversible logic synthesis begins from a given $n$-variable Boolean function, which can be irreversible. The first step is to convert it to a reversible Boolean function 
by adding distinguishing input bits initialized with a constant value. If these constant values are recovered after the reversible circuit execution, then these are known as \textit{ancilla} and otherwise, as \textit{garbage}.
\end{sloppypar}

\section{Related Work and Motivation}
\begin{figure}[hbt]
\caption{Reversible Circuit for $4mod5$}
\label{fig:4mod5}
\centering
\includegraphics[width=60mm]{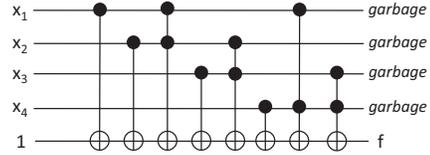}
\end{figure}

\begin{sloppypar}
Reversible logic synthesis methods can be broadly classified in four categories as following. A different and more detailed classification is presented in a recent survey 
of reversible logic synthesis methods~\cite{markov_survey}. \\
\textbf{Exact and Optimal methods:} In these methods, for small-scale reversible circuits the optimal implementation is found by making a step-by-step exhaustive 
enumeration or by formulating the reversible logic synthesis as a SAT problem~\cite{wille_sat} or reachability problem~\cite{reachability_hung}. Optimal implementation 
up to $4$-variable Boolean functions are known via exhaustive methods~\cite{shende_3var,golubitsky_4var} and up to few $6$-variable Boolean functions are known via 
SAT-based synthesis approach~\cite{wille_sat}. Exact methods perform well for small-scale circuits only since, reversible logic synthesis is shown to be an NP-hard problem~\cite{chatt_complexity}.\\
\textbf{Transformation-based methods~\cite{mmd,chandak_tx}:} These methods apply controlled transformations to map output Boolean functions to input Boolean functions. 
The method outlined in~\cite{chandak_tx} utilizes Boolean functions' nonlinearity measure and propose a column-wise synthesis approach, while~\cite{mmd} proceed 
row-wise in the boolean truth-table.\\
\textbf{Methods based on decision diagrams~\cite{wille_bdd,wille_addline,bdd_krishna}:} In these methods, each node of the decision diagram is converted to an 
equivalent reversible circuit structure. These methods reported excellent scaling for large Boolean functions, low QC at the cost of high number of garbage bits. 
A matrix-based representation is used to construct Quantm Multiple-valued Decision Diagram (QMDD) for reversible Boolean functions~\cite{qmdd}. This realizes ancilla-free 
reversible circuit and achieved QC comparable to transformation-based methods~\cite{wille_qmdd}.\\
\textbf{ESOP-based methods:} The advantage of ESOP formulation has been studied for classical logic synthesis, leading to state-of-the-art synthesis tools for 
obtaining an ESOP formulation with low number of cubes and literals~\cite{xorcism,rule_based_andxor}. For reversible logic synthesis, the ESOP formulation maps directly 
to the basic reversible logic gates. This synthesis method led to two different approaches so far. The first one, begins from a PPRM formulation, and heuristically 
searches for common kernels at each logic depth~\cite{esop_gupta}. In the other ESOP-based synthesis methods, each cube in the ESOP is converted into an equivalent 
Toffoli gate and all the cubes are \textit{xor}-ed in the dedicated output line, leading to a fixed garbage count. An example of straightforward application of this 
method is shown in the Figure~\ref{fig:4mod5}, where the reversible circuit for Grover's oracle ($4mod5$) is synthesized. The corresponding ANF is 
$1 \oplus x_1 \oplus x_2 \oplus x_1x_2 \oplus x_3 \oplus x_2x_3 \oplus x_4 \oplus x_1x_4 \oplus x_3x_4$.
\end{sloppypar}

\begin{sloppypar}
Realizing that the QC and gate count can be further reduced by sharing the common cubes, a shared-cube synthesis algorithm is proposed in~\cite{esop_fazel,esop_cube_nayeem}. 
Another method, proposed in~\cite{esop_tx}, identified two transformations to reduce the garbage and gate count. An evolutionary algorithm to generate reversible circuit 
using ESOP formulation is proposed in~\cite{esop_evo_wille}. 
\end{sloppypar}

\begin{sloppypar}
\textbf{Motivation:} Despite significant research in the reversible logic synthesis methods, the major focus has been so far towards synthesis with optimal 
garbage lines. While this is justified due to the difficulty of realizing a qubit in current quantum technologies, this also offers a restricted view of the 
synthesis performance. As it is shown in the methods based on the decision diagram~\cite{wille_bdd,wille_addline,wille_qmdd}, it is possible to achieve 
significantly less gate cost and QC than other state-of-the-art garbage-free synthesis method. Fundamentally, this reflects a trade-off, which is explored in this 
paper. The trade-off between ancilla and QC is presented in a recent paper~\cite{wille_tradeoff}. We propose similar approach, for ESOP-based reversible logic synthesis.
We show that, complementing the approach outlined in~\cite{wille_tradeoff}, the trade-off can be performed early in the logic synthesis flow. Several ESOP-based 
synthesis techniques~\cite{younes_bent,esop_cube_nayeem,esop_tx} focussed on reducing QC and gate count by common cube sharing and a set of 
transformations. In this paper, we attempt to generalize this and explore the link between ESOP minimization and reversible circuit synthesis performance. 
\end{sloppypar}

\begin{sloppypar}
A natural question at this point is whether it is possible to achieve ancilla-free reversible logic synthesis starting from an ESOP formulation, which represents 
one extreme point of the ancilla-cost trade-off. We propose a rule-based approach to achieve the same.
\end{sloppypar}

\begin{sloppypar}
In short, our contributions are twofold.
\begin{itemize}
 \item First, we show that diverse trade-off points between ancilla count and QC/Gate cost is achievable. We propose a tool-flow to perform the trade-off systematically.
 \item Second, we propose a rule-based, ancilla-free reversible logic synthesis technique starting from an ESOP formulation.
\end{itemize}

\end{sloppypar}

\begin{sloppypar}
The rest of this paper is organized as following. In the Section \ref{sec:synth_flow}, the overall synthesis flow is outlined. The ancilla-free ESOP-based reversible 
logic synthesis is detailed in the Section \ref{sec:ancilla_free}. The results are presented in the Section \ref{sec:results}. The paper is concluded with 
future directions in the Section \ref{sec:summary}.
\end{sloppypar}

\section{ESOP-based Synthesis Flow} \label{sec:synth_flow}
\begin{sloppypar}
The proposed ESOP-based synthesis flow starts from a canonical ANF representation. This is not an optimized ESOP representation like the one \cite{xorcism} adopted 
in other ESOP-based synthesis flows \cite{esop_gupta}\cite{esop_fazel}. However, the rationale of selecting this representation is that it offers an insight into the 
effect of the ESOP optimizations onto the quality of synthesis results. This is previously studied, in an evolutionary algorithm \cite{esop_evo_wille}.
\end{sloppypar}

\subsection{ANF Construction}
\begin{sloppypar}
Constructing an ANF from a Boolean truth table specification can be done using $O(n2^n)$ operations with standard algorithm. The input to the algorithm is the 
truth table $f = [f(0) f(1) f(2) \ldots f(2^n -1)]$, and the output is the coefficient vector of the ANF, represented as $C = [c_0 c_1 c_2 \ldots c_{2^n -1}]$. 
For an $n$-variable Boolean function, $C = fA_n$, where $A_n$ can be computed as following.
\end{sloppypar}
$
A_n = \begin{bmatrix}
       A_{n-1} & A_{n-1} \\
        0       & A_{n-1}
       \end{bmatrix}
, where A_0 = 1.
$
\vspace{0.2cm}
\begin{sloppypar}
Subsequently, the ANF is converted into an $n$-ary, directed acyclic graph (DAG) $G(V, E)$ for the optimization and mapping to reversible circuit. Note that, the ANF is 
considered as a starting point for the subsequent ESOP-based optimizations. Alternatively, this could be done with an ESOP-based form as available in~\cite{xorcism}. 
However, we started from a canonical representation to have better control at the optimizations of our choice. The DAG, that is generated from the ANF, is not the 
final circuit but, an intermediate representation for optimization and mapping. Here, $G(V, E)$ is defined as a set of nodes $V$ and a set of edges $E$, where $e_{i,j}$ 
indicates an edge between $v_i$ and $v_j$. The nodes can be of these types - \textit{t\_constant, t\_identifier, t\_root, t\_and} and \textit{t\_xor}. An exemplary DAG
for one output of $nth\_prime3\_inc$ circuit is shown in the following Figure~\ref{fig:nprime3}. The $root$ node is for combining all the different output functions 
of the reversible circuit. For every node, the $depth$ is appended as a suffix. The reversible Boolean specification of $nth\_prime3\_inc$ in permutation form is $\{0,2,3,5,7,1,4,6\}$~\cite{maslov_benchmark}.
\end{sloppypar}

\begin{figure}[hbt]
\caption{Graph-based Representation of $nth\_prime3\_inc$}
\label{fig:nprime3}
\centering
\includegraphics[width=90mm]{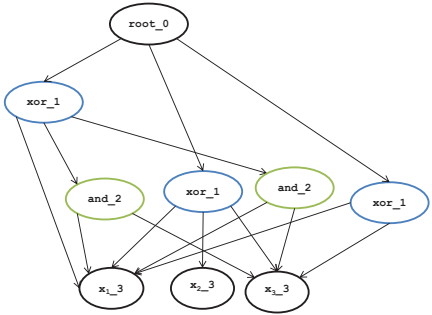}
\end{figure}

\subsection{Optimization Engines} \label{ssec:opt}
\begin{sloppypar}
In order to reduce the QC, gate cost and garbage count from the default implementation flow outlined in the previous section, several optimization engines are 
plugged in. These are described in the following subsections.
\end{sloppypar}

\begin{sloppypar}
\textbf{Kernel Extraction: } Kernel extraction from a polynomial expression is a well-studied problem in classical logic synthesis\cite{demicheli}. We start from an existing recursive 
kernel extraction algorithm (algorithm 8.3.4, \cite{demicheli}). The algorithm returns all possible kernels, $K(f)$ for a given function $f$. From those candidate 
kernels, the kernel with minimum remainder term ($k_l$) is chosen and termed as the divisor. After this initial kernel computation, this same process is repeated 
for the divisor, quotient and the remainder. Additionally, an user-defined threshold parameter is chosen to control the size of the selected Kernel. This ensures 
that the generated $and$-$xor$ graph is not skewed in nature and there is a fair distribution of edges across multiple depths. If no suitable kernel is found, the 
dividend is retained as it is with 2-level $and$-$xor$ graph formation.
\end{sloppypar}

\begin{algorithm}[htbp]
{\small
\caption{ExtractKernel}
\label{algoextractkernel}
\SetLine
\KwIn{$K(f)$}
\KwOut{$k_l$}
\ForAll{$k \in K(f)$} {
\If{$rem(f, k)$ $is$ $minimum$} {
\If{$k \rightarrow size() > threshold$}{
$return$ $k$\;
}}
}
$return$ $NULL$\;
}
\end{algorithm}

\begin{sloppypar}
\textbf{Common Cube Sharing: }As identified in several ESOP-based synthesis methods, sharing common cubes improves the synthesis quality signficantly. With the graph-based representation, 
it is convenient to perform the cube sharing optimization. Shareability of a pair of nodes is determined by the number of common children amongst them. If for 
either of the nodes, all the children are part of another node, then it is considered {\em shareable}. Otherwise, if the number of common children is more 
than half of the total children count of the two nodes, then the nodes are declared shareable. The nodes are shared across multiple depths of the $and$-$xor$ graph,
which again can be controlled as an input parameter.
\end{sloppypar}

\begin{algorithm}[htbp]
{\small
\caption{CommonCubeSharing}
\label{algocubesharing}
\SetLine
\For{$depth = depth_{max}-1 \to 1$} {
$depth_j = depth$\;
\ForAll{$node_i \in G(depth), node_j \in G(depth_j)$} {
\If{$shareable(node_i, node_j)$} {
$share(node_i, node_j)$\;
break\;
}
$depth_j = depth_j - 1$\;
}}}
\end{algorithm}

\begin{sloppypar}
\textbf{Parent Reduction Optimization: }A key observation from existing minimum garbage synthesis solutions~\cite{mmd} is that the final Toffoli network, when 
presented in the $and$-$xor$ graph form (see Figure~\ref{fig:nprime3}) allows for a perfect execution of the mapping algorithm to be presented later (Algorithm \ref{algofindtarget}). This formed the motivation of the parent reduction optimization. In this optimization, before every iteration of target node determination, 
the leaf node with minimum parents is identified as a candidate node. The number of parents for that particular node is minimized by applying the following 
expansion rules.
\end{sloppypar}

\begin{equation}
a = (a \oplus b) \oplus b
\end{equation}

\begin{equation}
a.b = ((a \oplus b)b) \oplus b
\label{rule2}
\end{equation}

\begin{sloppypar}
For both the rules, direct parents of node $a$ are avoided by adding an edge from the $a \oplus b$ node. This is beneficial only if there is an already and 
existing $a \oplus b$ parent node. This reduces the parent count of node $a$ at the cost of increased parent count of node $b$. This optimization is applied until 
no further reduction of node $a$ parent count is possible. Clearly, several more complex expansion rules can be formulated as well as a complex heuristic for 
determination of the target leaf node. However, that leads to a long runtime and therefore omitted. In principle, there exists one or more set of transformation 
rules to convert any given ESOP formulation to the ESOP representation realized by minimum garbage reversible logic synthesis methods~\cite{mmd}. For example, the 
following expansion rule is a more general form of the rule~\ref{rule2}, where $x$ is $0$.
\end{sloppypar}

\begin{equation}
a(b \oplus x) \oplus x = a(a \oplus b \oplus x) \oplus (a \oplus x)
\end{equation}

\begin{sloppypar}
Note that, these optimizations can be applied to any ESOP-based flow~\cite{esop_cube_nayeem,esop_evo_wille,esop_fazel,esop_gupta}. While, existing ESOP-based optimization approaches focus on reducing gate count or QC, we attempt to reduce line count here and in particular, to identify the trade-off between line count and QC.
\end{sloppypar}

\subsection{Mapping to Generalized Toffoli Network}
\begin{sloppypar}
Starting from the aforementioned graph representation, the mapping to Generalized Toffoli (GT) network is done by repeated determination of a target node 
($t\_node$) and then mapping it to the corresponding reversible circuit representation. Naturally, the target node determination forms the core algorithmic 
part, which is done heuristically as shown in the following Algorithm~\ref{algofindtarget}.
\end{sloppypar}

\begin{figure}[hbt]
\caption{Mapping of ESOP to Generalized Toffoli Circuit}
\label{fig:map}
\centering
\includegraphics[width=90mm]{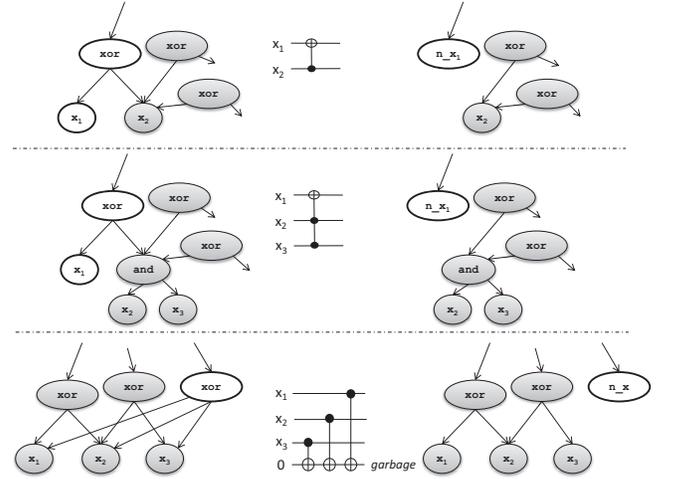}
\end{figure}

\begin{sloppypar}
The heuristic works in a greedy manner to minimize the garbage count. It traverses all the nodes in $depth$ above the leaf nodes. Whenever there is 
an available $xor$ node with a leaf node ($leaf\_single$) connected only to the $xor$ node, the $xor$ node is selected. All the other children of that 
node can perform the $xor$ operation and use the node $leaf\_single$ as target. If no such node is found, it proceeds towards all the $and$ nodes and looks 
for its parents, if it contains a similar $xor$ parent in $depth_{max}-2$ with single-parent leaf node. If no such node is found, the algorithm returns a node 
with maximum leaf node children. In case of a tie, the node with minimum parent is chosen. This leads to the creation of a garbage line. However, it creates 
the possibility to find single-parent in the next iteration. The callee mapping algorithm maps one node in one iteration, which is demonstrated the 
Figure~\ref{fig:map}. In the Figure~\ref{fig:map}, for every identified target node, the corresponding Toffoli circuit is shown in the middle and the modified 
graph is shown in the rightmost columns. Three different kinds of target identification, as presented in the Algorithm~\ref{algofindtarget}, are shown. It is 
straightforward to show that the algorithm~\ref{algofindtarget} always returns a valid node for mapping to the equivalent Toffoli circuit, ensuring that the 
mapping algorithm always terminates with a valid circuit synthesis.
\end{sloppypar}

\begin{algorithm}[htbp]
\SetLine
{\small
\caption{FindTarget}
\label{algofindtarget}
\KwIn{$G(V, E)$}
\KwOut{$t\_node$}
\ForAll{$node \in G(depth_{max}-1)$} {
\If{$node \rightarrow getType() == t\_xor$ $and$ $node \rightarrow containSingleChild()$} {
$return node$\;
}
\If{$node \rightarrow getType() == t\_and$ $and$ $node \rightarrow Parent() \rightarrow containSingleChild()$ $and$ $node \rightarrow Parent() \rightarrow getType() == t\_xor$} {
$return$ $node \rightarrow Parent()$\;
}
$return$ $nodeWithMaxChildMinParent()$\;
}
}
\end{algorithm}

\section{Ancilla-free, ESOP-based Reversible Logic Synthesis} \label{sec:ancilla_free}
\begin{sloppypar}
The proposed ancilla-free reversible logic synthesis algorithm treats the problem in a step-wise fashion starting from the output ESOP expressions. 
Each step applies a reversible transformation in order to reach a step, when all the functions are linear. From linear functions, the reversible circuit can be easily 
constructed. For example, the possible set of reversible transformations in a $3$-variable circuit consists of all possible $Tof_2$ and $Tof_3$ gates. We refer the 
application of these gates as transformation of types $T2$ and $T3$ respectively. All possible transformations are available as rules, one of which is chosen 
at each step by following a heuristic process.
\end{sloppypar}

\begin{sloppypar}
A transformation is adjudged suitable if it reduces the number of non-linear terms in the expression or reduces the number of literals in an already linear expression. 
Initially the non-linear terms are reduced using transformation of type $T2$. If it is no longer possible to apply $T2$ or two consecutive transformation are same, then 
transformation of type $T3$ are examined and a suitable transformation is chosen following a heuristic procedure. After all the output expressions have reduced to linear 
expressions, only $T2$ type transformation are applied which trivially reduce the expressions to single literal form. The key steps in the algorithm are checking the 
suitability of applying a specific transition with control and target variables, for which we developed heuristic techniques. These steps are explained in the following.
\end{sloppypar}

\begin{sloppypar}
\textbf{Suitability check of transformation type $T2$:} The algorithm checks for the occurrence of two non-linear terms in a single expression and then applies the $T2$ 
transformation with the control from the common variable and the target on one of the unique variables. The target is assigned arbitrarily. For example if $ab + bc$ is 
detected in an expression, the unique variables are $a$ and $c$. Either of these are arbitrarily chosen as the target variable with a $T2$ transformation. 
\end{sloppypar}

\begin{sloppypar}
\textbf{Suitability check of transformation type $T3$:} When $T2$ type transitions fail to further simplify the ESOP expressions, a target is arbitrarily fixed and 
appropriate control variables are searched for in the expressions where the target occurs. If the application of $T3$ leads to a decrease in the number of non-linear 
terms in the expressions then it is adjudged as suitable.
\end{sloppypar}

\begin{sloppypar}
\textbf{Terminating Condition:} After a cascade of changes the output expressions keep simplifying until the take the form of basic variables. This 
is when the algorithm terminates. It can be shown that either of these two previous transformations are always found to be suitable and the terminating 
condition is always achieved for $3$-variable functions.
\end{sloppypar}

\begin{sloppypar}
We explain the algorithmic flow with the following example. The three output expressions for the benchmark circuit $3\_17$ \cite{mmd_tcad} are as following.
\begin{equation}
f_1 = ac \oplus bc \oplus a \oplus c \oplus 1
\end{equation}
\begin{equation}
f_2 = a \oplus b \oplus c \oplus 1
\end{equation}
\begin{equation}
f_3 = ab \oplus bc \oplus b \oplus c \oplus 1
\end{equation}
\end{sloppypar}

\begin{sloppypar}
At the first stage, the presence of two nonlinear terms in $f_1$ is found, where $T2$ is suitable. The application of $T2$ with $b$ as control and $a$ as target line 
means that one needs to replace $a$ with $a \oplus b$ in all the expressions. The according changes in the aforementioned equations are as following.
\begin{equation}
f_1 = ac \oplus a \oplus b \oplus c \oplus 1
\end{equation}
\begin{equation}
f_2 = a \oplus c \oplus 1
\end{equation}
\begin{equation}
f_3 = ab \oplus bc \oplus c \oplus 1
\end{equation}
In the next step, two nonlinear terms in $f_3$ is found and again $T2$ with $c$ as control and $a$ as target line. By repeating these steps, one reaches the 
terminating condition.
\end{sloppypar}

\begin{algorithm}[htbp]
\SetLine
{\small
\caption{Ancilla-Free Reversible Logic Synthesis}
\label{algo_ancilla_free}
$f\_curr$ = $f\_out$\;
\While{$f\_curr$ $\neq$ $f\_in$} {
 \If{$f\_curr$ is nonlinear} {
  $found\_T2 = check\_T2(f\_curr, f\_mod)$\;
  \If{$found\_T2$ == $false$ OR $f\_last$ == $f\_mod$} {
    $apply\_T3(f\_curr, f\_mod)$\;
  } \Else {
    $f\_mod = apply\_T2(f\_curr)$\;
    $f\_last = f\_mod$\;
  }
 } \Else {
   $f\_mod = apply\_T2(f\_curr)$\;
 }
 $f\_curr$ = $f\_mod$\;
}
}
\end{algorithm}

\begin{sloppypar}
The algorithm is presented in form of pseudo-code in Algorithm~\ref{algo_ancilla_free}. The algorithm is run over the complete set of $3$-variable functions, of which the 
results are presented in the following section. For scaling the algorithm to more than $3$ variables, further heuristics with $Tof_4$ gates are applied. There, the goal 
is to first reduce the ESOP functions to consist of $2$-literal cubes only. Then onwards the above algorithm can be applied again. This gives rise to the possibility of 
reducing an $n$-variable reversible logic synthesis to $n-1$-variable form recursively. It must be noted that, $n$-literal cubes cannot be present for a $n$-variable 
reversible Boolean function to maintain balancedness property of the output functions.
\end{sloppypar}

\section{Results and Discussion} \label{sec:results}
\begin{sloppypar}
The abovementioned algorithms are implemented using C++, with around 3000 lines of code. The code is compiled with GCC version 4.6.6 and it is executed on an 
AMD Phenom\texttrademark II X6 1100T Processor running linux-based OS. In the following, we present a series of experiments to identify the advantage and 
disadvantage of the proposed methods vis-a-vis state-of-the-art reversible logic synthesis methods. The optimization kernels are controllable through command line 
parameters allowing various experiments. Though no specific optimization is turned on for increasing the count of Peres gates, those gates, when formed, are identified 
and accounted for in the QC and gate count computation. It must also be noted that final ordering of the output functions are chosen such as to minimize the QC \cite{wille_op_order}.
\end{sloppypar}

\subsection{QC-Ancilla Trade-off}
\begin{sloppypar}
For this experiment, the user-controllable parameters in the tool include the choice to run parent reduction (\textbf{P}), common cube sharing (\textbf{C}) 
optimization, the choice of threshold for kernel extraction (\textbf{K}) and selection of the size of the Toffoli gate (\textbf{T}). The last one is controlled 
when creating the initial $n$-ary ESOP graph. If the value of \textbf{C} is set to $true$, the selection of \textbf{T} has less influence. To show the trade-offs, 
two Boolean functions with significant relevance in cryptography, namely the S-Boxes for AES (8-variable) and PRESENT (4-variable) block ciphers are chosen. For 
the different values of parameters, corresponding gate cost, garbage count and QC values are presented in the Table~\ref{tab:tradeoff}. The best results for these 
functions obtained from~\cite{revkit} using BDD-based reversible logic synthesis is presented for comparison.
\end{sloppypar}

\begin{table}[htb]
  \caption{Trade-off Between QC and Garbage ($G^*$)}
  \label{tab:tradeoff}
  \centering
  \begin{tabular}{l|cccc|ccc}
    \hline

    \multirow{2}{*}{Function}      & \multicolumn{4}{c|}{this work} & \multicolumn{3}{c}{BDD\cite{revkit}} \\ \cline{2-8}
                                   & TCKP & QC     & Gates & $G^*$ & QC & Gates & $G^*$ \\ \hline
    \multirow{6}{*}{PRESENT SBox}  & 4000 & 148    &  24   &  3    & \multirow{6}{*}{86} &  \multirow{6}{*}{34}   & \multirow{6}{*}{7}\\
                                   & 3000 & 149    &  37   &  16   &    &       & \\
                                   & 3100 &  100   &  28   &  8    &    &       & \\
                                   & 3130 &  74    &  26   &  7    &    &       & \\
                                   & 3120 &  71    &  24   &  7    &    &       & \\
                                   & 3121 &  67    &  29   &  9    &    &       & \\ \hline
    \multirow{5}{*}{AES SBox}      & 9000 &  31161 &  1025 &  16    & \multirow{5}{*}{3277} &  \multirow{5}{*}{1005} & \multirow{5}{*}{200} \\ 
                                   & 6000 &  25196 &  1280 &  271  &      &       & \\
                                   & 3130 &  2281  &  825  &  276  &      &       & \\
                                   & 3150 &  2307  &  839  &  238  &      &       & \\
                                   & 3180 &  2420  &  896  &  188  &      &       & \\

  \hline
  \end{tabular}
\end{table}

\begin{sloppypar}
Table~\ref{tab:tradeoff} shows that the backend for mapping to Toffoli network can reduce the minimum garbage limit set by existing ESOP-based synthesis 
flows~\cite{esop_cube_nayeem}. For PRESENT S-Box, with maximum gate size being $Tof_4$, garbage count is $3$. With decreasing Toffoli size, QC decreases and 
gate count and garbage count increases, which is the effect of flattening. Enabling cube sharing reduces the gate count and garbage count to the original 
values and also significantly reduces the QC due to small-sized Toffoli gates. By enabling kernel extraction, QC can be further reduced at the expense of 
increased garbage count. Similar trend is found for the AES S-Box.
\end{sloppypar}

\begin{sloppypar}
For rest of the experiments, the focus is set on improving QC. We try to experiment with the proposed methodology and check if the QC and gate count
obtained by state-of-the-art synthesis methods can be reduced by compromising garbage count. To achieve that, common cube sharing is set to true, parent reduction optimization 
is set to false and size of the Toffoli gate is upper bounded by $3$. The kernel extraction threshold is initialized to $3$ and then, incremented in steps of $1$ until the 
QC values starts degrading. Typically, the best QC and garbage count is obtained below the threshold value of $7$. 
\end{sloppypar}

\subsection{Comparison with state-of-the-art ESOP-based Methods}
\begin{sloppypar}
It is important to study the benefits of the proposed trade-off capability of ESOP against existing ESOP-based methods reported in~\cite{esop_cube_nayeem,esop_gupta,esop_tx}. 
All of these methods begin with an optimized ESOP representation produced by~\cite{xorcism}. In~\cite{esop_cube_nayeem} several results for Boolean 
functions with large number of variables are reported. In~\cite{esop_gupta} detailed results for small Boolean functions are presented. However, both 
of these works have a minimum garbage count (which is greater than $0$). Reduced garbage count is reported in \cite{esop_tx}. The benchmarks presented 
in these papers are synthesized with our approach and presented in the Table~\ref{tab:esop}. The improved or matching QC values are marked with gray background.
\end{sloppypar}

\begin{table}[htb]
  \caption{Comparison with existing ESOP-based Methods}
  \label{tab:esop}
  \centering
  \begin{tabular}{l|ccc|ccc}
    \hline
    \multirow{2}{*}{Function}  & \multicolumn{3}{c|}{\cite{esop_cube_nayeem,esop_gupta,esop_tx}} & \multicolumn{3}{c}{this work} \\ \cline{2-7}
                    & QC  & Gates & $G^*$ & QC      & Gates & $G^*$ \\ \hline
    20f5 \cite{esop_gupta}       & 100 & 20  & 4   & 39\cellcolor[gray]{.8}  & 16\cellcolor[gray]{.8}  & 9 \\
    3\_17 \cite{esop_gupta}      & 14  & 6   & 0   & 19  & 11  & 4 \\
    4\_49 \cite{esop_gupta}      & 61  & 13  & 0   & 78  & 34  & 10 \\
    4mod5 \cite{esop_gupta}      & 13  & 5   & 3   & 9   & 5   & 4 \\
    5one013 \cite{esop_gupta}    & 95  & 19  & 4   & 48 \cellcolor[gray]{.8} & 17\cellcolor[gray]{.8}  & 10 \\
    5one245 \cite{esop_gupta}    & 104 & 20  & 4   & 44 \cellcolor[gray]{.8} & 13\cellcolor[gray]{.8}  & 8 \\
    5xp1 \cite{esop_tx}          & 695 & 45  & 11  & 755 & 271 & 97 \\
    6one135 \cite{esop_gupta}    & 5   & 5   & 5   & 5 \cellcolor[gray]{.8}  & 5\cellcolor[gray]{.8}   & 5 \\
    6one0246 \cite{esop_gupta}   & 6   & 6   & 5   & 6 \cellcolor[gray]{.8}  & 6\cellcolor[gray]{.8}   & 6 \\
    alu\cite{esop_gupta}         & 114 & 18  & 0   & 115 & 47  & 18 \\
    bw  \cite{esop_cube_nayeem}  & 676 & 322 & -   & 449 \cellcolor[gray]{.8}& 165\cellcolor[gray]{.8} & 82 \\
    decod24 \cite{esop_gupta}    & 31  & 11  & 0   & 77  & 25  & 8 \\
    f51m \cite{esop_tx}          & 501 & 32  & 8   & 228 \cellcolor[gray]{.8}& 77  & 43 \\
    graycode6 \cite{esop_gupta}  & 5   & 5   & 0   & 5 \cellcolor[gray]{.8}  & 5\cellcolor[gray]{.8}   & 0 \\
    majority3\cite{esop_gupta}   & 6   & 4   & 2   & 12  & 4   & 3 \\
    majority5\cite{esop_gupta}   & 104 & 16  & 4   & 45 \cellcolor[gray]{.8} & 13\cellcolor[gray]{.8}  & 9 \\
    mod5adder\cite{esop_gupta}   & 127 & 19  & 0   & 179 & 64  & 22 \\
    ham3\cite{esop_gupta}        & 9   & 5   & 0   & 7 \cellcolor[gray]{.8}  & 3\cellcolor[gray]{.8}   & 0 \\
    ham7\cite{esop_gupta}        & 68  & 24  & 0   & 76  & 32  & 11 \\
    hwb4\cite{esop_gupta}        & 35  & 15  & 0   & 64  & 24  & 8 \\
    rd32\cite{esop_gupta}        & 8   & 4   & 1   & 8 \cellcolor[gray]{.8}  & 2\cellcolor[gray]{.8}   & 2 \\
    rd53\cite{esop_tx}           & 91  & 17  & 7   & 44 \cellcolor[gray]{.8} & 11\cellcolor[gray]{.8}  & 11 \\
    rd73\cite{esop_tx}           & 525 & 43  & 9   & 94 \cellcolor[gray]{.8} & 29\cellcolor[gray]{.8}  & 20 \\
    sqr6\cite{esop_tx}           & 464 & 56  & 16  & 405 \cellcolor[gray]{.8}& 133 & 56 \\
    wim \cite{esop_tx}           & 83  & 23  & 10  & 106 & 38  & 20 \\
    xor5\cite{esop_tx}           & 4   & 4   & 4   & 4 \cellcolor[gray]{.8}  & 4\cellcolor[gray]{.8}   & 4 \\
    z4ml\cite{esop_tx}           & 329 & 33  & 8   & 478 & 170 & 61 \\
              \hline
  \end{tabular}
\end{table}

\begin{sloppypar}
Compared to the method proposed in this paper, all existing methods has considerably less execution time due to the fast heuristic ESOP 
minimization of EXORCISM-4~\cite{xorcism}. However, as can be observed from the results in the Table~\ref{tab:esop}, 
the solution is constrained by the output of EXORCISM-4, leading to poor results in some cases. This is expected, since none of the approaches 
\cite{esop_cube_nayeem,esop_gupta,esop_tx} experimented with the ESOP optimization parameters except for fixed transformations and cube sharing. 
Furthermore, in some cases the QC as well as the gate count is improved in our approach(e.g. \textit{$rd73$, $5one245$}). This reflects the importance 
of combined ESOP optimization and Toffoli-network mapping.
\end{sloppypar}

\subsection{Comparison with Decision Diagram-based Methods}
\begin{sloppypar}
While transformation-based methods represent the best results with minimum garbage for small functions, recent improvements in QMDD-based approach showed 
that the QC can be further lowered without increasing the garbage~\cite{wille_qmdd}. On the other hand, BDD-based synthesis achieve extremely small QC at the 
cost of increased garbage count.To do a fair benchmarking, our method is compared against the BDD-based method and QMDD-based method for the benchmarks reported 
in~\cite{wille_qmdd}. The results are presented in the Table~\ref{tab:all}. In the same mode as the previous experiments, our goal is set to reduce QC compared
to the reported results by increasing garbage count.
\end{sloppypar}

{\small
\begin{table*}[hbt]
  \caption{Comparison with Decision Diagram-based Methods}
  \label{tab:all}
  \centering
   \tabcolsep=0.11cm
  \begin{tabular}{lc|cccc|cccc|cccc}
    \hline
    \multirow{2}{*}{Function}  & \multirow{2}{*}{I/O} & \multicolumn{4}{c|}{BDD\cite{wille_bdd}} & \multicolumn{4}{c|}{QMDD\cite{wille_qmdd}} & \multicolumn{4}{c}{this work} \\ \cline{3-14}
                &       & QC  & Gates & Lines & Time(s) & QC & Gates & Lines & Time(s) & QC & Gates & Lines & Time(s) \\ \hline
adr4            & 8/5   & 237 & 93  & 33  & 0.19  &  5125    & 75  & 13  & 0.15  & 764   & 268 & 93  & 17.04 \\
clip            & 9/5   & 1196& 368 & 97  & 0.45  &  22495   & 232 & 14  & 0.48  & 1442  & 546 & 238 & 258.4 \\
cm42a           & 4/10  & 151 & 79  & 32  & 0.44  &  260     & 10  & 14  & 0.24  & 87\cellcolor[gray]{.8}    & 27  & 19  & 0.01 \\
cycle10\_2      & 12/12 & 202 & 78  & 39  & 0.07  &  10684   & 36  & 12  & 0.07  & 145\cellcolor[gray]{.8}   & 30  & 37  & 0.49 \\
dc1             & 4/7   & 193 & 77  & 28  & 0.03  &  426     & 25  & 11  & 0.02  & 130\cellcolor[gray]{.8}   & 54  & 21  & 0.01 \\
dc2             & 8/7   & 585 & 197 & 65  & 0.98  &  2974    & 60  & 15  & 0.56  & 1226  & 438 & 141 & 20.2 \\
dist            & 8/5   & 1023& 331 & 94  & 0.2   &  20624   & 241 & 13  & 0.31  & 1365  & 469 & 159 & 28.07 \\
max46           & 9/1   & 575 & 191 & 60  & 0.02  &  42248   & 51  & 10  & 0.04  & 729   & 261 & 83  & 64.28 \\
misex1          & 8/7   & 283 & 103 & 39  & 0.99  &  1000    & 35  & 15  & 0.74  & 1093  & 389 & 129 & 22.9 \\
plus127mod8192  & 13/13 & 98  & 54  & 25  & 0.17  &  9148    & 27  & 13  & 0.1   & 61\cellcolor[gray]{.8}    & 25  & 25  & 3218.85 \\
plus63mod4096   & 12/12 & 89  & 49  & 23  & 0.07  &  5413    & 26  & 12  & 0.04  & 58\cellcolor[gray]{.8}    & 22  & 24  & 1555.59 \\
radd            & 8/5   & 95  & 55  & 21  & 0.17  &  5125    & 75  & 13  & 0.17  & 849   & 297 & 109 & 25.78 \\
rd73            & 7/3   & 229 & 85  & 26  & 0.02  &  13858   & 147 & 10  & 0.04  & 94\cellcolor[gray]{.8}    & 29  & 20  & 0.08 \\
rd84            & 8/4   & 314 & 114 & 37  & 0.08  &  33900   & 278 & 12  & 0.16  & 208\cellcolor[gray]{.8}   & 58  & 30  & 0.41 \\
root            & 8/5   & 857 & 277 & 79  & 0.2   &  18497   & 204 & 13  & 0.27  & 1146  & 418 & 137 & 25.13 \\
sao2            & 10/4  & 725 & 237 & 76  & 0.35  &  9018    & 63  & 14  & 0.27  & 1476  & 504 & 192 & 13771 \\
sqrt8           & 8/4   & 259 & 95  & 31  & 0.08  &  3923    & 55  & 12  & 0.07  & 585   & 225 & 81  & 77.62 \\
squar5          & 5/8   & 267 & 99  & 36  & 0.18  &  704     & 30  & 13  & 0.13  & 151\cellcolor[gray]{.8}   & 43  & 28  & 0.01 \\
sqn             & 7/3   & 484 & 160 & 47  & 0.01  &  3507    & 50  & 10  & 0.03  & 485   & 177 & 61  & 1.15 \\
wim             & 4/7   & 134 & 62  & 25  & 0.04  &  239     & 13  & 11  & 0.03  & 106\cellcolor[gray]{.8}   & 38  & 20  & 0.01 \\
z4              & 7/4   & 187 & 75  & 26  & 0.04  &  3621    & 59  & 11  & 0.05  & 498   & 174 & 63  & 1.27 \\
\hline
  \end{tabular}
\end{table*}
}

\begin{sloppypar}
In the Table~\ref{tab:all}, instead of garbage count, the total lines are mentioned to synchronize with the results presented in~\cite{wille_qmdd}. 
The circuits for which QC is improved or matching compared to BDD-based flow are marked with gray scale. Compared to QMDD-based flow (as well as 
for~\cite{mmd}), the QC for all the circuits are improved. This is expected since the garbage count is compromised. It is interesting to note that, 
compared to BDD-based flow several circuits report improved performance in QC, gate count as well as garbage count for our ESOP-based method. This 
is possibly due to the difference in the Boolean structure optimization performed by BDD-based and ESOP-based method. The nature of Toffoli network 
mapping leads to high garbage count for both BDD and the ESOP-based method proposed in this paper.
\end{sloppypar}

\subsection{Comparison with Optimal and Known Best Results} \label{subsec:opt_results}
\begin{sloppypar}
It is interesting to verify the performance of the presented method against the known optimal results, in terms of gate-count and MCT library. 
The QC reductions considering Peres gates are also accounted for. The same library and QC model is also used in the performance assessment of our 
technique. We chose a set of benchmarks available in~\cite{shende_3var,golubitsky_4var,maslov_benchmark} and present the details in 
Table~\ref{tab:best}, where the improved or matching QC values are marked with gray shade. The results show that for Boolean functions with 
large variable count, the QC and gate count is improved while compromising the garbage count. Additionally for several small functions, we obtained 
overall improved or matching performance.
\end{sloppypar}

\begin{table}[hbt]
  \caption{Comparison with Optimal and Known Best Results}
  \label{tab:best}
  \centering
  \begin{tabular}{l|ccc|ccc}
    \hline
    \multirow{2}{*}{Function}  & \multicolumn{3}{c|}{Optimal/Known Best Result} & \multicolumn{3}{c}{this work} \\ \cline{2-7}
                        & QC  & Gates & Garbage & QC & Gates & Garbage \\ \hline
    2of5                & 75  & 17  & 5   & 39\cellcolor[gray]{.8}  & 16\cellcolor[gray]{.8}  & 9 \\
    3\_17               & 12  & 6   & 0   & 19  & 11  & 4 \\
    4\_49               & 28  & 14  & 0   & 78  & 34  & 10 \\
    4mod5               & 7   & 5   & 4   & 9   & 5   & 4 \\
    5mod5               & 76  & 10  & 5   & 36\cellcolor[gray]{.8}  & 16  & 8 \\
    6sym                & 62  & 20  & 9   & 71  & 20  & 12 \\
    9sym                & 94  & 28  & 11  & 488 & 183 & 81 \\
    cycle10\_2          & 1198& 19  & 0   & 145\cellcolor[gray]{.8} & 30  & 25 \\
    ham3                & 10  & 4   & 0   & 7\cellcolor[gray]{.8}   & 3\cellcolor[gray]{.8}   & 0 \\
    ham7                & 49  & 25  & 0   & 76  & 32  & 11 \\
    hwb4                & 19  & 13  & 0   & 64  & 24  & 8 \\
    hwb5                & 80  & 38  & 0   & 229 & 73  & 23 \\
    hwb6                & 107 & 47  & 0   & 446 & 172 & 58 \\
    hwb7                & 2611& 331 & 0   & 991\cellcolor[gray]{.8} & 364 & 120 \\
    hwb8                & 6940& 2710& 0   & 1846\cellcolor[gray]{.8}& 686\cellcolor[gray]{.8} & 234 \\
    nth\_prime\_3\_inc  & 6   & 4   & 0   & 6   & 4   & 0 \\
    nth\_prime\_4\_inc  & 26  & 14  & 0   & 70  & 26  & 8 \\
    nth\_prime\_5\_inc  & 80  & 36  & 0   & 176 & 68  & 22 \\
    nth\_prime\_6\_inc  & 667 & 55  & 0   & 383\cellcolor[gray]{.8} & 139 & 44 \\
    nth\_prime\_7\_inc  & 3172& 1427& 0   & 950\cellcolor[gray]{.8} & 354\cellcolor[gray]{.8} & 116 \\
    nth\_prime\_8\_inc  & 7618& 3346& 0   & 1741\cellcolor[gray]{.8}& 625\cellcolor[gray]{.8} & 218 \\
    rd32                & 8   & 4   & 2   & 8\cellcolor[gray]{.8}   & 2\cellcolor[gray]{.8}   & 2 \\
    rd73                & 64  & 20  & 7   & 94  & 29  & 17 \\
    rd84                & 98  & 28  & 11  & 208 & 58  & 26 \\
    xor5                & 4   & 4   & 4   & 4\cellcolor[gray]{.8}   & 4\cellcolor[gray]{.8}   & 4 \\
        \hline
  \end{tabular}
\end{table}

\begin{sloppypar}
For an exhaustive run on all $3$-variable Boolean function an average gate count of $7.6$, maximum gate size of $14$ and average garbage 
count of $2.3$ is obtained. In~\cite{esop_gupta}, garbage-free synthesis is done with an average gate count of $6.1$. 
The optimal gate count for NCT library is found as $5.87$~\cite{shende_3var}. Note that, in comparison with~\cite{esop_gupta}, our method is 
several orders of magnitude faster. While~\cite{esop_gupta} reports half a second synthesis time for each $3$-variable Boolean function, our 
entire exhaustive synthesis took only 11 seconds. 
\end{sloppypar}

\begin{sloppypar}
Though the ESOP-based methods reported in~\cite{esop_cube_nayeem,esop_fazel,esop_tx} would require $3$ garbage bits, the transformations 
mentioned in~\cite{esop_tx} resulted in significant improvement in gate cost and garbage count for exhaustive $3$-variable Boolean functions. 
There, an average gate count of $5.92$ and an average garbage count of $0.95$ is obtained. This shows that for small variable count, the optimized 
output produced by EXORCISM-4~\cite{xorcism} outweighs the optimizations presented in this paper.
\end{sloppypar}

\begin{sloppypar}
In the presented flow, it is observed that for Boolean functions with large variable count ($>10$), most of the processing time is spent in the 
optimization engines. The determination of target node and mapping to Toffoli circuit is extremely fast. When no optimization engine is run, the 
synthesis takes similar execution time compared to that reported in BDD-based methods. Presumably, the coupling of a fast and rule-based ESOP 
optimization flow such as \cite{xorcism} can improve the overall synthesis runtime. Interestingly, the generation of optimized ESOP in \cite{xorcism} 
uses a BDD-based representation.
\end{sloppypar}

\begin{sloppypar}
With a simple trade-off approach between QC and garbage count, we showed that improvement of QC is possible in a controlled manner. Moreover, better QC 
and gate count compared to state-of-the-art synthesis methods can be obtained, too. This findings can be crucial to develop reversible logic synthesis 
tools with more control over the synthesis outcome and to investigate further into the potential of ESOP-based synthesis methods.
\end{sloppypar}

\subsection{Ancilla-Free ESOP-based Reversible Logic Synthesis}
\begin{sloppypar}
For the ancilla-free approach outlined in the section~\ref{sec:ancilla_free}, we experimented with all $3$-variable Boolean functions and selected $4$-variable 
Boolean functions. For all the $3$-variable circuits our approach converged to an ancilla-free reversible circuit. For the $4$-variable Boolean functions, which 
we studied, all but one ($4bg15\_4$) function converged to ancilla-free result. Due to our focus on achieving ancilla-free synthesis from an ESOP formulation, 
the QC and gate counts fared poorly. For example, in case of $3$-variable circuits, our heuristics selected $Tof_2$ gates in a greedy manner. This potentially 
negelects globally optimal solutions with $Tof_3$ gates.
\end{sloppypar}

\begin{sloppypar}
For the $3$-variable Boolean functions, an average gate count of $9.28$ and an average QC of $17.14$ is obtained in our method. In contrast, the average 
gate count and QC values for optimal $3$-variable circuits are $5.87$ and $13.74$ respectively~\cite{shende_3var}. 
\end{sloppypar}



\section{Conclusion and Outlook} \label{sec:summary}
\begin{sloppypar}
In this paper, a parameterizable ESOP-based reversible logic synthesis flow is presented, which allows trade-off between QC and garbage count. 
The results are compared with state-of-the-art synthesis tools. It shows that significant benefits in performance can be obtained by tuning the 
optimizations and/or compromising the garbage count. Furthermore, ancilla-free ESOP-based reversible logic synthesis is proposed. The results, 
in terms of QC, is comparable to transformation-based reversible logic synthesis.\\
In future, further investigation will be done to appreciate the interplay between ESOP minimization and the quality of reversible circuit. 
Methods to relax the garbage count for better QC will be explored in the context of other reversible logic synthesis methods. Furthermore, the 
QC adopted for this paper needs to be updated with latest Clifford+T model of Quantum computing primitives.
\end{sloppypar}

\bibliographystyle{unsrt}

\begin{thebibliography}{9}

\bibitem{barenco}
{A. Barenco et al.}, ``{Elementary Gates for Quantum Computation},'' in {\em {Physical Review}}, 1995.

\bibitem{bennett}
{C. H. Bennett}, ``{Logical Reversibility of Computation},'' in {\em {IBM Journal of Research and Development}}, vol.~6, pp.~525--532, 1973.

\bibitem{chandak_tx}
{C. Chandak, A. Chattopadhyay, S. Majumder and S. Maitra}, ``{Analysis and Improvement of Transformation-Based Reversible Logic Synthesis},'' in {\em {Proceedings 
of the 2013 IEEE 43rd International Symposium on Multiple-Valued Logic (ISMVL '13)}}, IEEE Computer Society, Washington, DC, USA, 47-52. DOI=10.1109/ISMVL.2013.14 http://dx.doi.org/10.1109/ISMVL.2013.14 

\bibitem{chatt_complexity}
{A. Chattopadhyay, C. Chandak and K. Chakraborty}, ``{Complexity Analysis of Reversible Logic Synthesis},'' in {\em {arXiv, arXiv:1402.0491 [cs.ET]}}, February 2014. 

\bibitem{esop_bound}
{T. Sasao and P. Besslich}, ``{On the complexity of mod-2l sum PLA's}," in {\em {IEEE Transactions on Computers}}, vol.~39, no.~2, pp.~262--266, Feb 1990, doi: 10.1109/12.45212

\bibitem{demicheli}
{G. De Micheli}, ``{Synthesis and Optimization of Digital Circuits (1st ed.)}," McGraw-Hill Higher Education. ISBN:0070163332.

\bibitem{esop_evo_wille}
{R. Drechsler, A. Finder and R. Wille}, ``{Improving ESOP-based Synthesis of Reversible Logic Using Evolutionary Algorithms}," in { \em{Proceedings of the 2011 international conference on Applications of evolutionary computation - Volume Part II}}, pp.~151--161, 2011.

\bibitem{esop_fazel}
{K. Fazel, M. A. Thornton, and J. E. Rice}, ``{ESOP-based Toffoli Gate Cascade Generation}," in { \em{Proceedings of the IEEE Pacific Rim Conference on Communications, Computers and Signal Processing}}, 2007.

\bibitem{golubitsky_4var}
{O. Golubitsky, S. M. Falconer and D. Maslov}, ``{Synthesis of the Optimal 4-bit Reversible Circuits}," in { \em{Proceedings of DAC}}, pp.~653-656, 2010.

\bibitem{wille_sat}
{D. Grosse, R. Wille, G. W. Dueck and R. Drechsler},  ``{Exact Multiple-control Toffoli Network Synthesis with SAT Techniques}," in { \em{IEEE TCAD}},vol.~28, issue~5, May 2009.

\bibitem{esop_gupta}
{P. Gupta, A. Agrawal and N. K. Jha}, ``{An Algorithm for Synthesis of Reversible Logic Circuits}," in { \em{IEEE TCAD}}, vol.~25, no.~11, pp.~2317--2330, Nov. 2006.

\bibitem{reachability_hung}
{W. N. N. Hung, S. Xiaoyu, Y. Guowu, Y. Jin and M. Perkowski}, ``{Optimal Synthesis of Multiple Output Boolean Functions using a set of Quantum Gates by Symbolic Reachability analysis}," in {\em {IEEE TCAD}}, vol.~25, no.~9, pp.~1652--1663, Sept. 2006.

\bibitem{rule_based_andxor}
{D. Knysh and E. Dubrova}, ``{Rule-based Optimization of AND-XOR Expressions}," in { \em {Proceedings of the Reed-Muller Workshop}}, 2011.

\bibitem{bdd_krishna}
{M. Krishna and A. Chattopadhyay}, ``{Efficient Reversible Logic Synthesis via Isomorphic Subgraph Matching}," in {\em{Proceedings of the ISMVL}}, 2014.

\bibitem{mmd_tcad}
{D. Maslov, G. W. Dueck and D. M. Miller}, ``{Toffoli Network Synthesis with Templates}," in { \em{IEEE TCAD}}, vol.~24, issue~6, pp.~807--817, 2005.

\bibitem{mmd_fredkin}
{D. Maslov, G. W. Dueck and D. M. Miller}, ``{Synthesis of Fredkin-Toffoli Reversible Networks}," in { \em{IEEE TVLSI}}, vol.~13, issue~6, pp.~765--769, 2005.

\bibitem{mmd}
{D. M. Miller, D. Maslov and G. W. Dueck}, ``{A Transformation Based Algorithm for Reversible Logic Synthesis}," in { \em{Proceedings of DAC}}, 2003.

\bibitem{wille_addline}
{D. M. Miller, R. Wille and R. Drechsler}, ``{Reducing Reversible Circuit Cost by Adding Lines}," in {\em{Proceedings of the ISMVL}}, 2010.

\bibitem{wille_tradeoff}
{R. Wille, M. Soeken, D. M. Miller and R. Drechsler}, ``{Trading off circuit lines and gate costs in the synthesis of reversible logic}," in {\em{Integration, the VLSI Journal}}, Available online 21 September 2013, ISSN 0167-9260, http://dx.doi.org/10.1016/j.vlsi.2013.08.002.

\bibitem{qmdd}
{D. M. Miller and M. A. Thornton}, ``{QMDD: A Decision Diagram Structure for Reversible and Quantum Circuits}," in {\em{Proceedings of the 36th ISMVL}}, 2006.

\bibitem{miller_ismvl11}
{D. M. Miller, R. Wille, and Z. Sasanian}, ``{Elementary Quantum Gate Realizations for Multiple-Control Toffolli Gates}," in { \em{Proceedings of International Symposium on Multiple-Valued Logic (ISMVL)}}, pp.288-293, 2011.

\bibitem{xorcism}
{A. Mishchenko and M. Perkowski}, ``{Fast Heuristic Minimization of Exclusive-Sums-of-Products}," in { \em{Proceedings of the Reed-Muller Workshop}}, pp. 242-250, 2001.

\bibitem{maslov_benchmark}
{D. Maslov}, ``{Reversible Benchmarks},'' {\em{\url{http://webhome.cs.uvic.ca/~dmaslov}}}, last accessed May, 2014.

\bibitem{esop_cube_nayeem}
{N. M. Nayeem and J. E. Rice}, ``{Improved ESOP-based Synthesis of Reversible Logic}," in { \em{Proceedings of the 2011 Reed-Muller Workshop}}, 2011.

\bibitem{esop_tx}
{Y. Sanaee and G. W. Dueck}, ``{ESOP-Based Toffoli Network Generation with Transformations}," in { \em {Proceedings of the ISMVL}}, pp.~276-281, 2010.

\bibitem{markov_survey}
{M. Saeedi and I. L. Markov},``{Synthesis and Optimization of Reversible Circuits - A Survey}," in { \em{CoRR abs/1110.2574}}, {http://arxiv.org/abs/1110.2574}, 2011.

\bibitem{saaedi_cycle}
{M. Saeedi, M. S. Zamani, M. Sedighi and Z. Sasanian}, ``{Reversible Circuit Synthesis Using a Cycle-based Approach}," in { \em{J. Emerg. Technol. Comput. Syst}}, 2010.

\bibitem{shende_3var}
{V. V. Shende, A. K. Prasad, I. L. Markov and J. P. Hayes}, ``{Reversible Logic Circuit Synthesis}," in { \em{Proceedings of ICCAD}}, pp.~353-360, 2002.

\bibitem{revkit}
{M. Soeken, S. Frehse, R. Wille and R. Drechsler}, ``{RevKit: A Toolkit for Reversible Circuit Design}," in {\em{Workshop on Reversible Computation}}, 2010.

\bibitem{wille_qmdd}
{M. Soeken, R. Wille, C. Hilken, N. Przigoda and R. Drechsler}, ``{Synthesis of Reversible Circuits with Minimal Lines for Large Functions}," in { \em{Proceedings of the 17th ASP-DAC}}, 2012.

\bibitem{kerntopf_gen_peres}
{M. Szyprowski and P. Kerntopf}, ``{Low Quantum Cost Realization of Generalized Peres and Toffoli Gates with Multiple-Control Signals}," 
in{ \em{13th IEEE International Conference on Nanotechnology}}, 2013.

\bibitem{wille_bdd}
{R. Wille and R. Drechsler}, ``{BDD-based Synthesis of Reversible Logic for Large Functions}," in { \em{Proceedings of DAC}}, 2009.

\bibitem{wille_encoder}
{R. Wille, R. Drechsler, C. Osewold and A. Garcia-Ortiz}, ``{Automatic Design of Low-Power Encoders using Reversible Circuit Synthesis}," in { \em{Proceedings of the DATE}}, pp.1036-1041, 2012.

\bibitem{wille_op_order}
{R. Wille, D. Grosse, G. W. Dueck and R. Drechsler}, ``{Reversible Logic Synthesis with Output Permutation}," in { \em{Proceedings of the International Conference on VLSI Design}}, pp.189-194, 2009.

\bibitem{younes_bent}
{A. Younes},``{Synthesis and Optimization of Reversible Circuits for Homogeneous Boolean Functions}," in { \em{arXiv, arXiv:0710.0664v1}}, October 2007.

\end{thebibliography}

\end{document}